\documentclass[a4paper,11pt,
]{article}
\usepackage{amsmath,amssymb}
\usepackage[dvips]{color}
\usepackage{cite}
\usepackage{hangcaption}
\usepackage{amsmath,amssymb}
\usepackage[dvips]{color}
\usepackage{cite}
\usepackage{eepic}
\usepackage{epic}
\usepackage{hangcaption}
\usepackage{amsmath}
\usepackage{amssymb}
\usepackage{amscd}
\usepackage{amsthm}
\usepackage[dvipdfmx,%
 bookmarks=true,%
 bookmarksnumbered=true,%
 setpagesize=false,%
 pdfkeywords={TeX; dvipdfmx; hyperref; color;}]{hyperref}
\usepackage{hyperref}
\def\LL{\lq\lq}
\def\RR{\rq\rq}
\def\cases{\left\{\begin{array}{ll}}
\def\endcases{\end{array}\right.}
\def\roman{\rm}
\begin{document}
\setcounter{page}{1}
\vskip1.5cm
\begin{center}
{\LARGE \bf 
The Final Solutions of Monty Hall Problem and Three Prisoners Problem}
\vskip0.5cm
{\bf
\large
Shiro Ishikawa
}
\\
\vskip0.2cm
\rm
\it
Department of Mathematics, Faculty of Science and Technology,
Keio University, \\
3-14-1 Hiyoshi kohoku-ku, Yokohama, 223-8522 Japan.
\\
E-mail:
ishikawa@math.keio.ac.jp
\end{center}
\par
\rm
\vskip0.3cm
\par
\noindent
{\bf Abstract}
\normalsize
\par
\noindent
Recently we proposed the linguistic interpretation of quantum mechanics
(called quantum and classical measurement theory, or quantum language), which was characterized as a kind of metaphysical and linguistic turn of the Copenhagen interpretation.
This turn from physics to language does not only extend quantum theory to classical systems but also yield the quantum mechanical world view
(i.e., the philosophy of quantum mechanics, in other words, quantum philosophy).And we believe that this quantum language is the most powerful language to describe science. The purpose of this paper is to describe the Monty-Hall problem and the three prisoners problem in quantum language. We of course believe that our proposal is the final solutions of the two problems. Thus in this paper, we can answer the question: "Why have philosophers continued to stick to these problems?"And the readers will find that these problems are never elementary,
and  they can not be solved without the deep understanding
of "probability" and "dualism".

\rm

%
\par
\noindent

\par
%
%
%
%
\def\Cal{\cal}
\def\bigstimes{\text{\large $\: \boxtimes \,$}}

\par
\noindent
\section{
Introduction}
%
%
\subsection{
Monty Hall problem and Three prisoners problem
}
\rm
\par
According to 
ref. \cite{Ishi2},
we shall introduce the usual descriptions of the Monty Hall problem and the three prisoners problem as follows.
\par
\noindent
{\bf Problem 1}
\rm
[{}Monty Hall problem].
{
Suppose you are on a game show, and you are given
the choice of three doors
(i.e., \LL Door $A_1$\RR$\!\!\!,\;$ \LL Door $A_2$\RR$\!\!\!,\;$ \LL Door $A_3$\RR$\!\!)$.
Behind one door is a car, behind the others, goats.
You do not know what's behind the doors
\par
\noindent
However, you pick a door, say "Door $A_1$", and the host,
who knows what's behind the doors, opens another door,
say \LL Door $A_3$\RR$\!\!\!,\;$ which has a goat.
\par
\noindent
He says to you,
\LL Do you want to pick Door $A_2$?\RR$\;\;$
Is it to your advantage to switch your choice of doors?
\par
\noindent
\vskip0.3cm
\par
\noindent
\unitlength=0.26mm
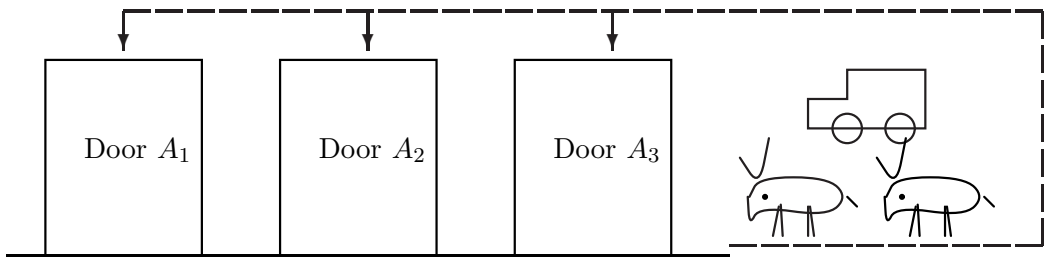
\begin{picture}(500,150)
\thicklines
\put(430,55)
{{
\drawline[-15](-40,-30)(120,-30)(120,90)(-350,90)
\put(-350,90){\vector(0,-1){20}}
\put(-225,90){\vector(0,-1){20}}
\put(-100,90){\vector(0,-1){20}}
\path(0,30)(60,30)(60,60)(20,60)(20,45)(0,45)(0,30)
\put(20,30){\circle{15}}
\put(47,30){\circle{15}}
}}
\put(400,20)
{{
\spline(0,30)(5,40)(40,40)(50,30)(40,20)(5,25)(0,15)(-1,30)
\spline(-5,50)(5,35)(10,60)
\path(15,25)(12,10)
\path(16,26)(17,10)
\path(30,25)(30,10)
\path(31,25)(33,10)
\put(8,30){\circle*{2}}
\path(50,30)(55,25)
}}
\put(470,20)
{{
\spline(0,30)(5,40)(40,40)(50,30)(40,20)(5,25)(0,15)(-1,30)
\spline(-5,50)(5,35)(10,60)
\path(15,25)(12,10)
\path(16,26)(17,10)
\path(30,25)(30,10)
\path(31,25)(33,10)
\put(8,30){\circle*{2}}
\path(50,30)(55,25)
}}

\thicklines
\put(20,20){\line(1,0){370}}
\put(40,20){
\path(0,0)(0,100)(80,100)(80,0)
\put(20,50){Door $A_1$}
}
\put(160,20){
\path(0,0)(0,100)(80,100)(80,0)
\put(20,50){Door $A_2$}
}
\put(280,20){
\path(0,0)(0,100)(80,100)(80,0)
\put(20,50){Door $A_3$}
}
\end{picture}

\par
\noindent
{\bf Problem 2}
\rm
[{}Three prisoners problem].
Three prisoners, $A_1$, $A_2$, and $A_3$ were in jail.
They knew that one of them was to be set free and
the other two were to be executed.
They did not know who was the one to be spared,
but the emperor did know.
$A_1$ said to the emperor, 
{\LL}I already know that at least one the other two prisoners will be executed, so if you tell me the name of one who will be executed, you won't have given me any information about my own execution\RR.$\;\;$
After some thinking, the emperor said,
\LL $A_3$ will be executed.\RR$\;\;$
Thereupon $A_1$ felt happier because
his chance had increased from $\frac{1}{3(= {\roman Num}[\{A_1,A_2,A_3 \}])}$ to 
$\frac{1}{2(= {\roman Num}[\{ A_1,A_2 \}])}$.
This prisoner $A_1$'s happiness may or may not be reasonable?
\par
\noindent
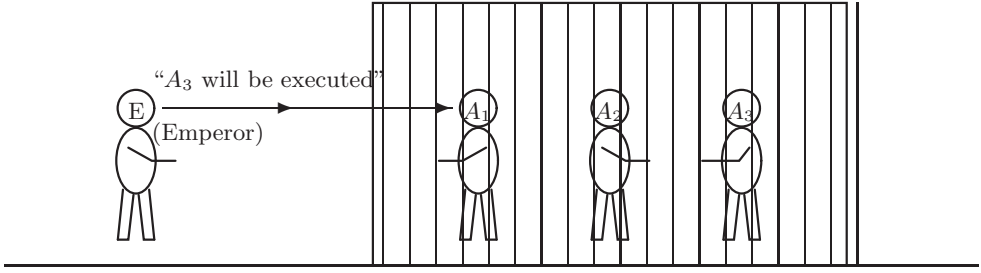
\unitlength=0.35mm
\begin{picture}(500,130)
\thicklines
\put(20,0)
{{{
\put(70,20)
{
\put(0,60){\circle{14}}
\put(0,40){\ellipse{15}{25}}
\path(-3,45)(6,40)(15,40)
\put(-3,56){\footnotesize E}
\path(-7,10)(-5,29)
\path(5,29)(7,10)
\path(-7,10)(-3,10)(-1,27)
\path(1,27)(3,10)(7,10)
}
\put(200,20)
{{
{
\put(0,60){\circle{14}}
\put(0,40){\ellipse{15}{25}}
\path(3,45)(-6,40)(-15,40)
\put(-6,56){\footnotesize $A_1$}
\path(-7,10)(-5,29)
\path(5,29)(7,10)
\path(-7,10)(-3,10)(-1,27)
\path(1,27)(3,10)(7,10)
}
\put(50,0)
{
\put(0,60){\circle{14}}
\put(0,40){\ellipse{15}{25}}
\path(-3,45)(6,40)(15,40)
\put(-6,56){\footnotesize $A_2$}
\path(-7,10)(-5,29)
\path(5,29)(7,10)
\path(-7,10)(-3,10)(-1,27)
\path(1,27)(3,10)(7,10)
}
\put(100,0)
{
\put(0,60){\circle{14}}
\put(0,40){\ellipse{15}{25}}
\path(3,45)(-,40)(-15,40)
\put(-6,56){\footnotesize $A_3$}
\path(-7,10)(-5,29)
\path(5,29)(7,10)
\path(-7,10)(-3,10)(-1,27)
\path(1,27)(3,10)(7,10)
}
}}
\thicklines
\put(20,20){\line(1,0){370}}
\put(160,20){
\path(0,0)(0,100)(180,100)(180,0)
}
\linethickness{0.15mm}
\put(164,20)
{
\multiput(0,0)(10,0){19}{\line(0,1){100}}
}
\put(70,20)
{
\put(10,60){\vector(1,0){50}}
\put(60,60){\vector(1,0){60}}
\put(6,68){\footnotesize \LL $A_3$ will be executed\RR}
\put(6,48){\footnotesize (Emperor)}
}
}}}
\end{picture}

The purpose of this paper
is to clarify
Problem 1 (Monty Hall problem)
and Problem 2 (three prisoners problem )
as follows.
\begin{itemize}
\item[(A1)]Problem 1 (Monty Hall problem) is solvable, but Problem 2 (Three prisoners problem) is not well posed. In this sense, Problem 1 and Problem 2  are not equivalent.
This is the direct consequence of Fisher's maximal likelihood method mentioned in Section 2.
\item[(A2)]Also, there are two ways that the probabilistic property is introduced to both problems as follows:
\begin{itemize}
\item[(A2$_1$)]
in Problem 1,
one (discussed in Section 4) is that the host casts the dice, and another (discussed in Section 6) is that
you cast the dice.
\item[(A2$_2$)]
in Problem 2,
one (discussed in Section 4) is that the emperor casts the dice, and another (discussed in Section 6) is that
three prisons cast the dice.
\end{itemize}
In the case of each,
the former solution is due to Bayes' method ( mentioned in Section 2).
And the latter solution is due to
the principle is equal probabilities ( mentioned in Section 5).
And, after all, we can conclude,
under the situation (A2), 
that
Problem 1 and Problem 2 are equivalent.
\end{itemize}

%
%
%
%

\rm
The above will be shown in terms of quantum language
(=measurement theory).
And therefore, we expect the readers to find that
quantum language is superior to Kolmogorov's probability theory
{\cite{Kolm}}.
%


%
%


\par
\noindent
\subsection{
Overview: Measurement Theory (= Quantum Language) }
\rm
\par

\rm
\par
\par
\rm
As emphasized in
refs.
{}{\cite{Ishi4,Ishi5}},
measurement theory
(or in short, MT)
is,
by a linguistic turn of quantum
mechanics
(cf. {\bf Figure 1}:\textcircled{\footnotesize 7} later), constructed
as the scientific
theory
formulated
in a certain 
{}{$C^*$}-algebra ${\cal A}$
(i.e.,
a norm closed subalgebra
in the operator algebra $B(H)$
composed of all bounded operators on a Hilbert space $H$,
{\rm cf.} {}{\cite{Murp, Neum}}
).
MT is composed of
two theories
(i.e.,
pure measurement theory
(or, in short, PMT]
and
statistical measurement theory
(or, in short, SMT).
That is, it has the following structure:
\par
\rm
\par
\begin{itemize}
\item[(B)]
$
\underset{\text{\footnotesize }}{
\text{
MT (measurement theory = quantum language)
}
}
$
\\
$=\cases
\text{(B1)}:
\underset{\text{\scriptsize }}{\text{[PMT
]}}
=
\displaystyle{
{
\mathop{\mbox{[(pure) measurement]}}_{\text{\scriptsize (Axiom$^{\rm P}$ 1) }}
}
}
+
\displaystyle{
\mathop{
\mbox{
[causality]
}
}_{
{
\mbox{
\scriptsize
(Axiom 2)
}
}
}
}
\\
\\
\text{(B2)}
:
\underset{\text{\scriptsize }}{\text{[SMT
]}}
=
\displaystyle{
{
\mathop{\mbox{[(statistical) measurement]}}_{\text
{\scriptsize (Axiom$^{\roman S}$ 1) }}
}
}
\!
+
\!
\displaystyle{
\mathop{
\mbox{
[causality]
}
}_{
{
\mbox{
\scriptsize
(Axiom 2)
}
}
}
}
\endcases
$
\end{itemize}
\par
\noindent
where
Axiom 2 is common in PMT and SMT.
For completeness, note that measurement theory (B)
(i.e.,
(B1) and (B2))
is not physics but 
a kind of language
based on
{\lq\lq}the quantum mechanical world view{\rq\rq}.
As seen in
{}{\cite{Ishi6}},
note that
MT gives a foundation to statistics.
That is,
roughly speaking,
\begin{itemize}
\item[(C1)]
it may be understandable
to
consider that
PMT and SMT
is related to
Fisher's statistics
and
Bayesian statistics
respectively.
\end{itemize}


%

When ${\cal A}=B_c(H)$,
the ${C^*}$-algebra composed
of all compact operators on a Hilbert space $H$,
the (B) is called {quantum measurement theory}
(or,
quantum system theory),
which can be regarded as
the linguistic aspect of quantum mechanics.
Also, when ${\cal A}$ is commutative
$\big($
that is, 
when ${\cal A}$ is characterized by $C_0(\Omega)$,
the $C^*$-algebra composed of all continuous 
complex-valued functions vanishing at infinity
on a locally compact Hausdorff space $\Omega$
({\rm cf.} {}{\cite{Murp}})$\big)$,
the (B) is called {classical measurement theory}.
Thus, we have the following classification:
\begin{itemize}
\item[(C2)]
$
\quad
\underset{\text{\scriptsize }}{\text{MT}}
$
$\left\{\begin{array}{ll}
\text{quantum MT$\quad$(when ${\cal A}=B_c (H)$)}
%
\\
\\
\text{classical MT
$\quad$
(when ${\cal A}=C_0(\Omega)$)}
\end{array}\right.
$
\end{itemize}

%

\par
\noindent
Also,
for the position of
MT
in science,
see {\bf Figure 1},
which was precisely explained in {}{\cite{Ishi5, Ishi8}}.


\begin{center}
\begin{picture}(410,170)
{
\put(10,70){
{
\put(0,-3){
$\!\!
\underset{{\text{\footnotesize Alistotle}}}{\underset{{\text{\footnotesize Plato}}}{\overset{\text{\footnotesize Parmenides}}{{\overset{\text{\footnotesize Socrates}}{
{\fbox{\shortstack[l]{Greek\\ 
{\footnotesize philosophy}}}}
}
}
}
}
}
$
}
}
\put(51,-3){
\rm
$\xrightarrow[\text{\footnotesize sticism}]{\text{\footnotesize Schola-} }$
$\!\! \textcircled{\scriptsize 1}$
}
\put(93,7){
{\line(0,1){34}}
}
\put(93,-7){
{\line(0,-1){47}}
}
}
\put(100,70){
$
\begin{array}{l}
\!\!\!
{\; \xrightarrow[]{ \; 
\quad
}}
\overset{\text{\scriptsize (monism)}}{\underset{\text{\scriptsize (realism)}}
{\fbox{\text{Newton}}}}
{
\overset{\textcircled{\scriptsize 2}}{{\line(1,0){17}}}
}
\begin{array}{llll}
\!\!
\rightarrow
{\fbox{\shortstack[l]{relativity \\ theory}}}
{\xrightarrow[]{\qquad \quad \;\;\;}
}{\textcircled{\scriptsize 3}}
\\
\\
\!\!
\rightarrow
{\fbox{\shortstack[l]{quantum \\mechanics}}}
{
\xrightarrow[\qquad \quad \;\; ]{}
}\textcircled{\scriptsize 4}
\end{array}
\\
\\
\!\! \xrightarrow[]{{
\quad}}
\overset{\text{\scriptsize (dualism)}}{
\underset{\text{\scriptsize (idealism)}}{\fbox{
{\shortstack[l]{Descartes\\ Rock,... \\Kant}}
}}
}
{\xrightarrow[]{\textcircled{\scriptsize 6}
}}
\!
\overset{\text{\scriptsize (linguistic view)}}{\fbox{
\shortstack[l]{language \\ philosophy}
}}
\!\! \xrightarrow[{
}
{
\text{\footnotesize
zation
}}
]{{
{
}
{\text{\footnotesize
quanti-}
}
}}\textcircled{\scriptsize 8}
\end{array}
$
}
\put(300,86){
{\put(-40,0){\drawline(0,2)(0,-30)}}
{\put(-40,-32){\text{$\xrightarrow[]{\; \text{\footnotesize language}}$}}}
{\put(6,-32){\textcircled{\scriptsize 7}}}
}
\put(190,80){\line(0,1){46}}
\put(302,110){
$
\left.\begin{array}{llll}
\; 
\\
\; 
\\
\; 
\\
\;
\end{array}\right\}
\xrightarrow[]{\textcircled{\scriptsize 5}}
{\!\!\!
\overset{\text{\scriptsize (unsolved)}}{
\underset{\text{\scriptsize (quantum phys.)}}{
\fbox{\shortstack[l]{theory of \\ everything}}
}
}
}
$
}
\put(302,20){
$
\left.\begin{array}{lllll}
\; 
\\
\; 
\\
\; 
\\
\;
\\
\;
\end{array}\right\}
{\xrightarrow[]{\textcircled{\scriptsize 10}}}
\overset{\text{\scriptsize (=MT)}}{
\underset{\text{\scriptsize (language)}}{
\fbox{\shortstack[l]{\color{red}quantum\\ \color{red}language}}
}
}
$
}
\put(100,-70){
{\bf 
\hypertarget{Figure 1}{Figure 1}: The history of the world-view
}
}
\put(65,-32){
}
{
\thicklines
\color{red}
\dashline[50]{4}(287,-47)(270,-47)(270,70)(420,70)(420,-47)(380,-47)}
\thicklines
{\put(175,-16)
{{
{
\fbox{\shortstack[l]{ statistics \\ system theory}}
}
}$\xrightarrow[]{\qquad \;}$\textcircled{\scriptsize 9}
}
}
{
{
\put(288,-50){\color{red}
\bf
$\;\;$
linguistic view
}
\color{black}
}
{
\put(200,155){\color{blue}
\bf
$\;\;$
realistic view
}
\color{black}
}
}
}
{
\color{blue}
$\!\!\!\!\!\!\!\!${\dashline[50]{4}(190,160)(110,160)(110,74)(420,74)(420,160)(290,160)}
\color{black}
}
\end{picture}
\vskip1.8cm
\end{center}

\par
\vskip0.3cm
\par
\noindent

\section{
Classical Measurement Theory
(Axioms
and
Interpretation)}

\subsection{
Mathematical Preparations
}


\par
\noindent
\par
Since our concern is concentrated to the Monty Hall problem and
three prisoners problem,
we devote ourselves to
classical MT in (C2).
\par
Throughout this paper,
we assume that
$\Omega$ is
a
compact Hausdorff space.
Thus, we can put
$C_0(\Omega) =$
$C(\Omega)$,
which is
defined by
a Banach space
(or precisely, a commutative $C^*$-algebra
)
composed of all continuous 
complex-valued functions
on a compact Hausdorff space $\Omega$,
where
its norm $\|f\|_{C(\Omega)}$
is defined by
$\max_{\omega \in \Omega}|f(\omega)|$.

Let
${C(\Omega)}^*$ be the
dual Banach space of
${C(\Omega)}$.
That is,
$ {C(\Omega)}^* $
$ {=}  $
$ \{ \rho \; | \; \rho$
is a continuous linear functional on ${C(\Omega)}$
$\}$,
and
the norm $\| \rho \|_{ {C(\Omega)}^* } $
is defined by
$ \sup \{ | \rho ({}f{}) |  \:{}: \; f \in {C(\Omega)}
\text{ such that }\| f \|_{{C(\Omega)}} \le 1 \}$.
The bi-linear functional
$\rho(f)$
is
also denoted by
${}_{{C(\Omega)}^*}
\langle \rho, f \rangle_{C(\Omega)}$,
or in short
$
\langle \rho, f \rangle$.
Define the
\it
mixed state
$\rho \;(\in{C(\Omega)}^*)$
\rm
such that
$\| \rho \|_{{C(\Omega)}^* } =1$
and
$
\rho ({}f) \ge 0
\text{ 
for all }f\in {C(\Omega)}
\text{ such that }
f \ge 0$.
And put
\begin{align*} {\frak S}^m  ({}{C(\Omega)}^*{})
{=}
\{ \rho \in {C(\Omega)}^*  \; | \;
\rho
\text{ is a mixed state}
\}.
\end{align*}
%
\rm
Also, for each $\omega \in \Omega$,
define the {\it pure state}
$\delta_\omega$
$(\in
{\frak S}^m  ({}{C(\Omega)}^*{})
)$
such that
${}_{{C(\Omega)}^*}
\langle \delta_\omega, f \rangle_{C(\Omega)}$
$=$
$f(\omega)$
$(\forall f \in C(\Omega ))$.
And put
\begin{align*} {\frak S}^p ({}{C(\Omega)}^*{})
{=}
\{ \delta_\omega \in {C(\Omega)}^*  \; | \;
\delta_\omega 
\text{ is a pure state}
\},
\end{align*}
which is called a
\it
state space.
\rm
%
\rm
Note,
by the Riesz theorem
({\rm cf}.
{}{\cite{Yosi}}
), that
$C(\Omega )^*$
$=$
${\cal M}(\Omega )
$
$\equiv$
$\{
\rho \;|\;
\rho
$
is a signed measure on $\Omega$
$
\}$
and
${\frak S}^m(C(\Omega )^*)$
$=$
${\cal M}_{+1}^m(\Omega )
$
$\equiv$
$\{
\rho \;|\;
\rho
$
is a measure on $\Omega$
such that
$\rho(\Omega)=1$
$
\}$.
Also,
it is clear that
$ {\frak S}^p  ({}{C(\Omega)}^*{})$
$=$
$\{ \delta_{\omega_0} \;|\; \delta_{\omega_0}$ is a point measure at
${\omega_0}
\in \Omega
\}$,
where
$ 
\int_\Omega f(\omega) \delta_{\omega_0} (d \omega )$
$=$
$f({\omega_0})$
$
(\forall f
$
$
\in C(\Omega))$.
This implies that
the state space
$ {\frak S}^p  ({}{C(\Omega)}^*{})$
can be also identified with
$\Omega$
(called a {\it spectrum space}
or simply,
{\it spectrum})
such as
\begin{align}
\underset{\text{\scriptsize (state space)}}{{\frak S}^p  ({}{C(\Omega)}^*{})}
\ni \delta_{\omega} \leftrightarrow {\omega} \in 
\underset{\text{\scriptsize (spectrum)}}{\Omega}
\label{eq1}
\end{align}
Also, note that
${C(\Omega)}$
is unital,
i.e.,
it
has the identity $I$
(or precisely,
$I_{C(\Omega)}$),
since we assume that
$\Omega$ is compact.

According to the noted idea ({\rm cf.} {}{\cite{ Davi}})
in quantum mechanics,
an {\it observable}
${\mathsf O}{\; :=}(X, {\cal F},$
$F)$ in 
${{C(\Omega)}}$
is defined as follows:
\par
\par
\begin{itemize}
\item[(D$_1$)]
[Field]
$X$ is a set,
${\cal F}
(\subseteq 2^X $,
the power set of $X$)
is a field of $X$,
that is,
{\lq\lq}$\Xi_1, \Xi_2 \in {\cal F}\Rightarrow \Xi_1 \cup \Xi_2 \in {\cal F}${\rq\rq},
{\lq\lq}$\Xi  \in {\cal F}\Rightarrow X \setminus \Xi \in {\cal F}\;${\rq\rq}.
\item[(D$_2$)]
[Additivity]
$F$ is a mapping from ${\cal F}$ to ${{C(\Omega)}}$ 
satisfying:
(a):
for every $\Xi \in {\cal F}$, $F(\Xi)$ is a non-negative element in 
${{C(\Omega)}}$
such that $0 \le F(\Xi) $
$\le I$, 
(b):
$F(\emptyset) = 0$ and 
$F(X) = I$,
where
$0$ and $I$ is the $0$-element and the identity
in ${C(\Omega)}$
respectively.
(c):
for any $\Xi_1$,
$\Xi_2$
$\in {\cal F}$
such that
$\Xi_1 \cap \Xi_2 = \emptyset$,
it holds that
$
F(\Xi_1 \cup \Xi_2 )
$
$
=
$
$
F(\Xi_1  )
+
F( \Xi_2 )$.
\end{itemize}
\par
\noindent
%
%
\par
\noindent
For the more precise argument
(such as countably additivity, etc.),
see {}{\cite{Ishi6}}.
\vskip0.3cm
\par
\par

\par

\par
\noindent
\subsection{
Classical PMT
in (B1)
}
\rm
\par
In this section
we shall explain classical PMT
in
(A$_1$).
\par
\rm
With any {\it system} $S$, a commutative $C^*$-algebra 
${C(\Omega)}$
can be associated in which the 
measurement theory (B) of that system can be formulated.
A {\it state} of the system $S$ is represented by an element
${\delta_\omega} (\in {\frak S}^p  ({}{C(\Omega)}^*{}))$
and an {\it observable} is represented by an observable 
${\mathsf{O}}{\; :=} (X, {\cal F}, F)$ in ${{C(\Omega)}}$.
Also, the {\it measurement of the observable ${\mathsf{O}}$ for the system 
$S$ with the state ${\delta_\omega}$}
is denoted by 
${\mathsf{M}}_{{{C(\Omega)}}} ({\mathsf{O}}, S_{[{\delta_\omega}]})$
$\big($
or more precisely,
${\mathsf{M}}_{C(\Omega)} ({\mathsf{O}}{\; :=} (X, {\cal F}, F), S_{[{\delta_\omega}]})$
$\big)$.
An observer can obtain a measured value $x $
($\in X$) by the measurement 
${\mathsf{M}}_{C(\Omega)} ({\mathsf{O}}, S_{[{\delta_\omega}]})$.
\par
\noindent
\par
The Axiom$^{\rm P}$ 1 presented below is 
a kind of mathematical generalization of Born's probabilistic interpretation of quantum mechanics.
And thus, it is a statement without reality.
\par
\noindent
{\bf{Axiom$^{\rm P}$ 1\;\;
\rm
$[$Classical PMT$]$}}.
\it
The probability that a measured value $x$
$( \in X)$ obtained by the measurement 
${\mathsf{M}}_{{{C(\Omega)}}} ({\mathsf{O}}$
${ :=} (X, {\cal F}, F),$
{}{$ S_{[{\delta_{\omega_0}}]})$}
%
belongs to a set 
$\Xi (\in {\cal F})$ is given by
$
[F(\Xi)](\omega_0)
$.
\rm

\par
\par
\vskip0.2cm
\par

\par
Next, we explain Axiom 2 in (B).
Let $(T,\le)$ be a tree, i.e., a partial ordered 
set such that {\lq\lq$t_1 \le t_3$ and $t_2 \le t_3$\rq\rq} implies {\lq\lq$t_1 \le t_2$ or $t_2 \le t_1$\rq\rq}\!.
In this paper,
we assume that
$T$ is finite.
Assume that
there exists an element $t_0 \in T$,
called the {\it root} of $T$,
such that
$t_0 \le t$ ($\forall t \in T$) holds.
Put $T^2_\le = \{ (t_1,t_2) \in T^2{}\;|\; t_1 \le t_2 \}$.
The family
$\{ \Phi_{t_1,t_2}{}: $
${C(\Omega_{t_2})} \to {C(\Omega_{t_1})} \}_{(t_1,t_2) \in T^2_\le}$
is called a {\it causal relation}
({\it due to the Heisenberg picture}),
\rm
if it satisfies the following conditions {}{(E$_1$) and 
(E$_2$)}.
\begin{itemize}
\item[{\rm (E$_1$)}]
With each
$t \in T$,
a $C^*$-algebra ${C(\Omega_{t})}$
is associated.
\item[{\rm (E$_2$)}]
For every $(t_1,t_2) \in T_{\le}^2$, a Markov operator 
$\Phi_{t_1,t_2}{}: {C(\Omega_{t_2})} \to {C(\Omega_{t_1})}$ 
is defined
(i.e.,
$\Phi_{t_1,t_2} \ge 0$,
$\Phi_{t_1,t_2}(I_{{C(\Omega_{t_2})}})$
$
=
$
$
I_{{C(\Omega_{t_1})}}$
).
And it satisfies that
$\Phi_{t_1,t_2} \Phi_{t_2,t_3} = \Phi_{t_1,t_3}$ 
holds for any $(t_1,t_2)$, $(t_2,t_3) \in T_\le^2$.
\end{itemize}
\noindent
The family of dual operators
$\{ \Phi_{t_1,t_2}^*{}: $
$
{\frak S}^m  ({C(\Omega_{t_1})}^*)
\to {\frak S}^m  ({C(\Omega_{t_2})}^*)
\}_{(t_1,t_2) \in T^2_\le}$
is called a
{
\it
dual causal relation}
({\it
due to the Schr\"{o}dinger picture}).
When
$ \Phi_{t_1,t_2}^*{}$
$
(
{\frak S}^p  ({C(\Omega_{t_1})}^*)
$
$\subseteq
$
$
{\frak S}^p  ({C(\Omega_{t_2})}^*)
$
holds for any
$
{(t_1,t_2) \in T^2_\le}$,
the causal relation is said to be
deterministic.


\par
\par
\rm
Here, Axiom 2 in the measurement theory (B) is presented
as follows:
\rm
\par
\noindent
{\bf{Axiom 2}
\rm[Causality]}.
\it
The causality is represented by
a causal relation 
$\{ \Phi_{t_1,t_2}{}: $
${C(\Omega_{t_2})} \to {C(\Omega_{t_1})} \}_{(t_1,t_2) \in T^2_\le}$.

\rm
\par
For the further argument
(i.e.,
the $W^*$-algebraic formulation) of measurement theory,
see
Appendix in {}{\cite{Ishi4}}.
\noindent
%

\par
\noindent
\par
\noindent

%

\noindent

%
%

\subsection{
Classical SMT
in (B2)
}


\rm


\rm
It is usual to consider that
we do not know the state
$\delta_{\omega_0}$
when
we take a measurement
${\mathsf{M}}_{{{C(\Omega)}}} ({\mathsf{O}}, S_{[\delta_{\omega_0}]})$.
That is because
we usually take a measurement ${\mathsf{M}}_{{{C(\Omega)}}} ({\mathsf{O}},
S_{[\delta_{\omega_0}]})$
in order to know the state $\delta_{\omega_0}$.
Thus,
when we want to emphasize that
we do not know the the state $\delta_{\omega_0}$,
${\mathsf{M}}_{{{C(\Omega)}}} ({\mathsf{O}}, S_{[\delta_{\omega_0}]})$
is denoted by
${\mathsf{M}}_{{{C(\Omega)}}} ({\mathsf{O}}, S_{[\ast]})$.
Also,
when we know the distribution $\nu_0$
$( \in {\cal M}_{+1}^m(\Omega) ={\frak S}^m({C(\Omega)}^*) )$
of the unknown state
$\delta_{\omega_0}$,
the
${\mathsf{M}}_{{{C(\Omega)}}} ({\mathsf{O}}, S_{[\delta_{\omega_0}]})$
is denoted by
${\mathsf{M}}_{{{C(\Omega)}}} ({\mathsf{O}},$
$ S_{[\ast]}
( \nu_0 ) )$.
%

\par
\vskip0.3cm

\par
\par
The Axiom$^{\rm S}$ 1 presented below is 
a kind of mathematical generalization of 
Axiom$^{\rm P}$ 1.

\par

\par
\noindent
{\bf{Axiom$^{\rm S}$\;1\;
\rm
\;[Classical SMT]}}.
\it
The probability that a measured value $x$
$( \in X)$ obtained by the measurement 
${\mathsf{M}}_{{{C(\Omega)}}} ({\mathsf{O}}$
${ :=} (X, {\cal F}, F),$
{}{$ S_{[\ast]}( \nu_0 ) )$}
%
belongs to a set 
$\Xi (\in {\cal F})$ is given by
$
\nu_0 ( F(\Xi) )
$
$($
$=
{}_{{{C(\Omega)}^*}}\langle
\nu_0,
F(\Xi)
\rangle_{{C(\Omega)}}$
$)$.
\rm


\par
\noindent
{\it Remark 1}.
Note that
two statistical measurements
${\mathsf{M}}_{{{C(\Omega)}}} ({\mathsf{O}},$
{}{$ S_{[\delta_{\omega_1}]}( \nu_0 ) )$}
and
${\mathsf{M}}_{{{C(\Omega)}}} ({\mathsf{O}},$
{}{$ S_{[\delta_{\omega_2}]}( \nu_0 ) )$}
can not be distinguished
before
measurements.
In this sense,
we consider that,
even if $\omega_1 \not= \omega_2$,
we can assume that
\begin{align}
{\mathsf{M}}_{{{C(\Omega)}}} ({\mathsf{O}},
{}{ S_{[\delta_{\omega_1}]}( \nu_0 ) )}
=
{\mathsf{M}}_{{{C(\Omega)}}} ({\mathsf{O}},
{}{ S_{[\ast]}( \nu_0 ) )}
=
{\mathsf{M}}_{{{C(\Omega)}}} ({\mathsf{O}},
{}{ S_{[\delta_{\omega_2}]}( \nu_0 ) )}.
\label{eq2}
\end{align}

\par
\noindent
\vskip0.2cm
\par

\subsection{
Linguistic Interpretation
}

\par
Next,
we have to
answer how to use the above axioms
as follows.
That is, we present the following 
linguistic interpretation
(F)
[=(F$_1$)--(F$_3$)],
which is characterized as a kind of linguistic turn
of so-called Copenhagen interpretation
({\rm cf.}
{}{\cite{Ishi4,Ishi5}}
).

\par
\noindent
That is,
we propose:
\begin{itemize}
\item[(F$_1$)]
Consider the dualism composed of {\lq\lq}observer{\rq\rq} and {\lq\lq}system( =measuring object){\rq\rq}
such as
\par
\noindent
\vskip0.5cm
\noindent
\begin{center}
\unitlength=0.5mm
\begin{picture}(200,72)(15,0)
\put(-8,0)
{
\allinethickness{0.2mm}
\drawline[-40](80,0)(80,62)(30,62)(30,0)
\drawline[-40](130,0)(130,62)(175,62)(175,0)
\allinethickness{0.5mm}
\path(20,0)(175,0)
%
\put(14,-5){
\put(37,50){$\bullet$}
}
\put(50,25){\ellipse{17}{25}}
\put(50,44){\ellipse{10}{13}}
\put(0,44){\put(43,30){\sf \footnotesize{observer}}
\put(42,25){\scriptsize{(I(=mind))}}
}
\put(7,7){\path(46,27)(55,20)(58,20)}
\path(48,13)(47,0)(49,0)(50,13)
\path(51,13)(52,0)(54,0)(53,13)
\put(0,26){
\put(142,48){\sf \footnotesize system}
\put(143,43){\scriptsize (matter)}
}
\path(152,0)(152,20)(165,20)(150,50)(135,20)(148,20)(148,0)
\put(10,0){}
\allinethickness{0.2mm}
\put(0,-5){
\put(130,39){\vector(-1,0){60}}
\put(70,43){\vector(1,0){60}}
\put(92,56){\sf \scriptsize \fbox{observable}}
\put(58,50){\sf \scriptsize }
\put(57,53){\sf \scriptsize \fbox{\shortstack[l]{measured \\ value}}}
\put(80,44){\scriptsize \textcircled{\scriptsize a}interfere}
\put(80,33){\scriptsize \textcircled{\scriptsize b}perceive a reaction}
\put(130,56){\sf \scriptsize \fbox{state}}
}
}
\put(30,-15){\bf
\hypertarget{fig2}{\bf Figure 2}. 
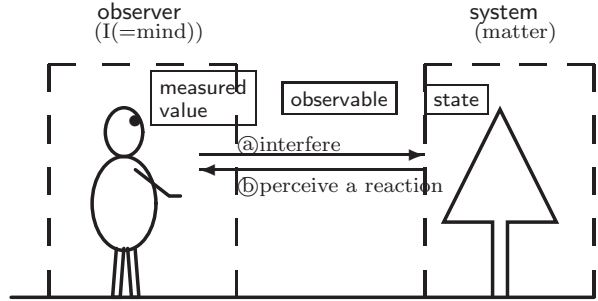
Descartes' figure
in MT
}
\end{picture}
\end{center}
\vskip1.0cm
And therefore,
{\lq\lq}observer{\rq\rq} and {\lq\lq}system{\rq\rq}
must be absolutely separated.
\item[(F$_2$)]
Only one measurement is permitted.
And thus,
the state after a measurement
is meaningless
$\;$
since it 
can not be measured any longer.
Also, the causality should be assumed only in the side of system,
however,
a state never moves.
Thus,
the Heisenberg picture should be adopted.
\item[(F$_3$)]
Also, the observer
does not have
the space-time.
Thus, 
the question:
{\lq\lq}When and where is a measured value obtained?{\rq\rq}
is out of measurement theory,
\end{itemize}
\par
\noindent
and so on.
This interpretation is,
of course,
common to
both PMT
and
SMT.

%
%

\par
\noindent
\vskip0.2cm
\par

\subsection{
Preliminary Fundamental Theorems
}

We have the following two fundamental theorems
in measurement theory:

\par
\noindent
{\bf Theorem 1}
[Fisher's maximum likelihood method
({\rm cf}. {}{\cite{Ishi6}})].
Assume that
a measured value obtained
by a measurement
${\mathsf M}_{C(\Omega)}({\mathsf O}:=(X,{\cal F}, F) , S_{[*]})$ belongs to
$\Xi \;(\in {\cal F} )$.
Then,
there is a reason to infer that
the unknown state
$[\ast]$
is equal to 
$\delta_{\omega_0}$,
where
$\omega_0 \;(\in \Omega )$
is defined by
\begin{align*}
[{F}(\Xi )](\omega_0)
= \max_{\omega \in \Omega  
}
[{F}(\Xi )](\omega).
\end{align*}

\par
\noindent
{\bf Theorem 2}
[Bayes' method
({\rm cf}. {}{\cite{Ishi6}})].
Assume that
a measured value obtained
by a statistical measurement
${\mathsf M}_{C(\Omega)}({\mathsf O}:=(X,{\cal F}, F) ,$
$S_{[*]}( \nu_0 ))$ belongs to
$\Xi \;(\in {\cal F} )$.
Then,
there is a reason to infer that
the posterior state
(i.e.,
the mixed state
after the measurement
)
is equal to $\nu_{\roman post}$,
which
is defined by
\begin{align*}
\nu_{\roman post} (D)
=
\frac{\int_D [F(\Xi)](\omega ) \nu_0(d \omega) }{\int_\Omega [F(\Xi)](\omega ) \nu_0(d \omega) }
\\
\quad
(\forall D \in {\cal B}_\Omega;
\text{Borel field}).
\end{align*}

%
The above two theorems are, of course, the most fundamental
in statistics.
Thus, we believe in
{\bf Figure 1},
i.e.,
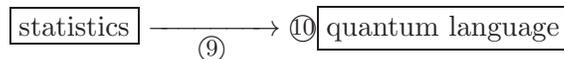
$$
\fbox{\mbox{statistics}}
\xrightarrow[\textcircled{\footnotesize 9}]{\qquad \qquad}\textcircled{\footnotesize 10}
\fbox{\mbox{quantum language}}
$$
%
%

\par
\noindent
\section{
The First Answer to Monty Hall Problem
\textcolor{blue}{[resp. Three prisoners problem]}
by Fisher's method}
\par

%
%
%

\par
\noindent
\par
In this section,
we present the first answer to
Problem 1
(Monty-Hall problem)
\textcolor{blue}{[resp. Problem 2
(Three prisoners problem)]}
in classical PMT.
The two will be simultaneously solved as follows.
The spirit of dualism (in Figure 2) urges us to declare that
\begin{itemize}
\item[(G)]
"observer$\;\; \approx\;\; $you" and "system$\;\; \approx\;\; $three doors" in Problem 1
\\
\color{blue}
[resp. "observer$\;\; \approx\;\; $prisoner $A_1$" and "system$\;\; \approx \;\;$emperor's mind" in Problem 2]
\color{black}
\end{itemize}
\par
\noindent
Put
$\Omega = \{ \omega_{1} , \omega_{2} , \omega_{3} \}$
with the discrete topology.
Assume that
each state
$\delta_{\omega_{{m}}}
(\in 
{\frak S}^p (C(\Omega)^* ))$
means
\begin{align}
&
\delta_{\omega_{{m}}}
\Leftrightarrow
\text{
the state that the car is }
\text{behind the door $A_m$}
\nonumber
\\
\textcolor{blue}{[\mbox{resp.}}
\;\;\;
&
\mbox{
\textcolor{blue}{
$
\delta_{\omega_{{m}}}
\Leftrightarrow
$
}}
\mbox{
\textcolor{blue}{
the state that the prisoner $A_m$ will be free
]
}
}
\nonumber
\\
&
\qquad
\qquad
(m=1,2,3 )
\label{eq3} \end{align}
Define the observable
${\mathsf O}_1$
$\equiv$
$({}\{ 1, 2,3 \}, 2^{\{1, 2 ,3\}}, F_1)$
in $C({}\Omega{})$
such that
\begin{align}
& [F_1({}\{ 1 \}{})](\omega_1{})= 0.0,\qquad
[F_1({}\{ 2 \}{})](\omega_1{})= 0.5,
\qquad 
[F_1({}\{ 3 \}{})](\omega_1{})= 0.5,
\nonumber
\\
&
[F_1({}\{ 1 \}{})](\omega_2{})= 0.0,
\qquad
[F_1({}\{ 2 \}{})](\omega_2{})= 0.0, \qquad 
[F_1({}\{ 3 \}{})](\omega_2{})= 1.0,
\nonumber
\\
& [F_1({}\{ 1 \}{})](\omega_3{})= 0.0,\qquad
[F_1({}\{ 2 \}{})](\omega_3{})= 1.0, \qquad 
[F_1({}\{ 3 \}{})](\omega_3{})= 0.0,
\label{eq4} \end{align}
where
it is also possible to assume that
$F_1({}\{ 2 \}{})(\omega_1{})=\alpha$,
$F_1({}\{ 3 \}{})(\omega_1{}) =1- \alpha$
$
(0 < \alpha < 1)$.
Thus we have a measurement
${\mathsf M}_{C({}\Omega{})} ({}{\mathsf O}_1, S_{[{}\ast{}]})$,
which should be regarded as
the measurement theoretical representation of
the measurement
that
\it
you say "Door $A_1$"
\rm
\textcolor{blue}{
[resp.
\it
"Prisoner $A_1$" asks to the emperor
\rm
].}
Here, we assume that
\begin{itemize}
\item[a)]
{\lq\lq}measured value $1$ is obtained by
the measurement
${\mathsf M}_{C({}\Omega{})} ({}{\mathsf O}_1, S_{[{}\ast{}]})${\rq\rq}
\\
$
\Leftrightarrow \text{The host says {\lq\lq}Door $A_1$
has a goat{\rq\rq}}$
\\
\color{blue}
\text{[resp.
$\Leftrightarrow$
the emperor says {\lq\lq}Prisoner $A_1$
will be executed{\rq\rq}
]}
\color{black}
\item[b)]
{\lq\lq}measured value $2$ is obtained
by
the measurement
${\mathsf M}_{C({}\Omega{})} ({}{\mathsf O}_1, S_{[{}\ast{}]})$
{\rq\rq}
\\
$
\Leftrightarrow \text{The host says {\lq\lq}Door $A_2$
has a goat{\rq\rq}}$
\\ \color{blue}
\text{[resp.
$\Leftrightarrow$
the emperor says {\lq\lq}Prisoner $A_2$
will be executed{\rq\rq}
]}
\color{black}
\item[c)]
{\lq\lq}measured value $3$ is obtained
by
the measurement
${\mathsf M}_{C({}\Omega{})} ({}{\mathsf O}_1, S_{[{}\ast{}]})$
{\rq\rq}
\\
$
\Leftrightarrow \text{The host says {\lq\lq}Door $A_3$
has a goat{\rq\rq}}$
\\
\color{blue}
\text{[resp.
$\Leftrightarrow$
the emperor says {\lq\lq}Prisoner $A_3$
will be executed{\rq\rq}
]}
\color{black}
\end{itemize}
\par

\par
\noindent
Recall that,
in 
Problem 1
(Monty-Hall problem)
\textcolor{blue}{[resp. Problem 2
(Three prisoners problem)]}, 
the host said
{\lq\lq}Door 3 has a goat{\rq\rq}
\textcolor{blue}{[resp. the emperor said
{\lq\lq}Prisoner $A_3$ wil be executed{\rq\rq}]}
This implies that
you \textcolor{blue}{[resp. {\lq\lq}Prisoner $A_1$]}
get the measured value {\lq\lq}3{\rq\rq}
by the measurement
${\mathsf M}_{C({}\Omega{})} ({}{\mathsf O}_1, S_{[\ast]}{})$.
Note that
\begin{align}
&
[F_1({}\{3\}{})] ({}\omega_2{})
=
1.0
=
\max \{ 0.5, \; \; 1.0 , \; \; 0.0 \}
\nonumber
\\
&
=
\max \{
[F_1({}\{3\}{})] ({}\omega_1{}),
[F_1({}\{3 \}{}){}]({}\omega_2{}),
[F_1({}\{3 \}{})] ({}\omega_3{})
\},
\label{eq5} \end{align}
Therefore,
Theorem 1
(Fisher's maximum likelihood method)
says that
\begin{itemize}
\item[(H1)]
In
Problem 1
(Monty-Hall problem),
there is a reason to infer that
$[\ast]$
$=$
$\delta_{\omega_2}$.
Thus,
you should
switch to Door $A_2$.
\color{blue}
\item[(H2)]
\color{blue}
In
Problem 2
(Three prisoners problem),
there is a reason to infer that
$[\ast]$
$=$
$\delta_{\omega_2}$.
However,
there is no reasonable answer for the question:
whether
Prisoner $A_1$'s happiness increases.
That is, Problem 2 is not a well-posed problem.
\color{black}
\end{itemize}

\par
\noindent
\section{
The Second Answer to Monty Hall Problem
\textcolor{blue}{[resp. Three prisoners problem]}
by Bayes' method}
\par
In order to use Bayes' method, shall modify Problem 1(Monty Hall problem) and
Problem 2(Three prisoners problem) as follows.

\subsection{
Problems 1$'$ and 2$'$
(
Monty Hall Problem
\textcolor{blue}{[resp. Three prisoners problem]}
)
}

\par
\noindent
{\bf Problem 1$'$}
\rm
[{}Monty Hall problem; the host casts the dice].
{
Suppose you are on a game show, and you are given
the choice of three doors
(i.e., \LL Door $A_1$\RR$\!\!\!,\;$ \LL Door $A_2$\RR$\!\!\!,\;$ \LL Door $A_3$\RR$\!\!)$.
Behind one door is a car, behind the others, goats.
You do not know what's behind the doors.
\par
\noindent
However, you pick a door, say "Door $A_1$", and the host,
who knows what's behind the doors, opens another door,
say \LL Door $A_3$\RR$\!\!\!,\;$ which has a goat.
And he adds that
\begin{itemize}
\rm
\item[{}{($\sharp_1$)}]
\it
the car was set
behind the door
decided by the cast of the (distorted) dice.
That is,
the host set the car
behind Door $A_m$
with probability
$p_m$
(where
$p_1 + p_2 + p_3 =1$, $ 0 \le p_1 , p_2 , p_3 \le 1 $
$)$.
\end{itemize}
He says to you,
\LL Do you want to pick Door $A_2$?\RR$\;\;$
Is it to your advantage to switch your choice of doors?
%
%
%
%
\par
\noindent
\vskip0.3cm
\par
\noindent
\unitlength=0.26mm
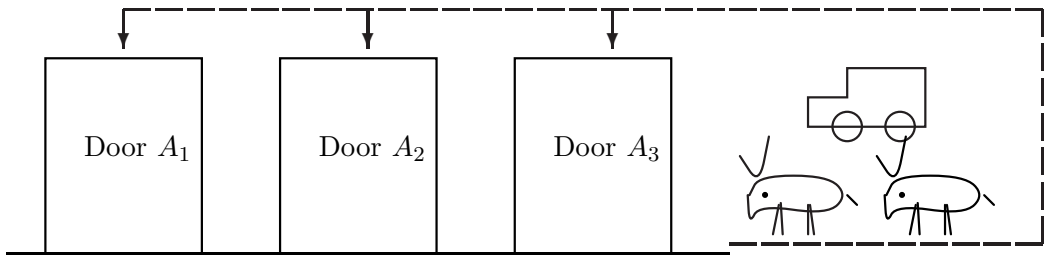
\begin{picture}(500,150)
\thicklines
\put(430,55)
{{
\drawline[-15](-40,-30)(120,-30)(120,90)(-350,90)
\put(-350,90){\vector(0,-1){20}}
\put(-225,90){\vector(0,-1){20}}
\put(-100,90){\vector(0,-1){20}}
\path(0,30)(60,30)(60,60)(20,60)(20,45)(0,45)(0,30)
\put(20,30){\circle{15}}
\put(47,30){\circle{15}}
}}
\put(400,20)
{{
\spline(0,30)(5,40)(40,40)(50,30)(40,20)(5,25)(0,15)(-1,30)
\spline(-5,50)(5,35)(10,60)
\path(15,25)(12,10)
\path(16,26)(17,10)
\path(30,25)(30,10)
\path(31,25)(33,10)
\put(8,30){\circle*{2}}
\path(50,30)(55,25)
}}
\put(470,20)
{{
\spline(0,30)(5,40)(40,40)(50,30)(40,20)(5,25)(0,15)(-1,30)
\spline(-5,50)(5,35)(10,60)
\path(15,25)(12,10)
\path(16,26)(17,10)
\path(30,25)(30,10)
\path(31,25)(33,10)
\put(8,30){\circle*{2}}
\path(50,30)(55,25)
}}

\thicklines
\put(20,20){\line(1,0){370}}
\put(40,20){
\path(0,0)(0,100)(80,100)(80,0)
\put(20,50){Door $A_1$}
}
\put(160,20){
\path(0,0)(0,100)(80,100)(80,0)
\put(20,50){Door $A_2$}
}
\put(280,20){
\path(0,0)(0,100)(80,100)(80,0)
\put(20,50){Door $A_3$}
}
\end{picture}

\par
\noindent
{\bf Problem 2$'$}
\rm
[{}Three prisoners problem; the emperor casts the dice].
Three prisoners, $A_1$, $A_2$, and $A_3$ were in jail.
They knew that one of them was to be set free and
the other two were to be executed.
They did not know who was the one to be spared, but they know that
\begin{itemize}
\rm
\item[{}{($\sharp_2$)}]
\it
the one to be spared
was
decided by the cast of the (distorted) dice.
That is,
Prisoner $A_m$ is to be spared
with probability
$p_m$
(where
$p_1 + p_2 + p_3 =1$, $ 0 \le p_1 , p_2 , p_3 \le 1 $
$)$.
\end{itemize}
but the emperor did know the one to be spared.
$A_1$ said to the emperor, 
{\LL}I already know that at least one the other two prisoners will be executed, so if you tell me the name of one who will be executed, you won't have given me any information about my own execution\RR.$\;\;$
After some thinking, the emperor said,
\LL $A_3$ will be executed.\RR$\;\;$
Thereupon $A_1$ felt happier because
his chance had increased from $\frac{1}{3(= {\roman Num}[\{A_1,A_2,A_3 \}])}$ to 
$\frac{1}{2(= {\roman Num}[\{ A_1,A_2 \}])}$.
This prisoner $A_1$'s happiness may or may not be reasonable?
\par
\noindent
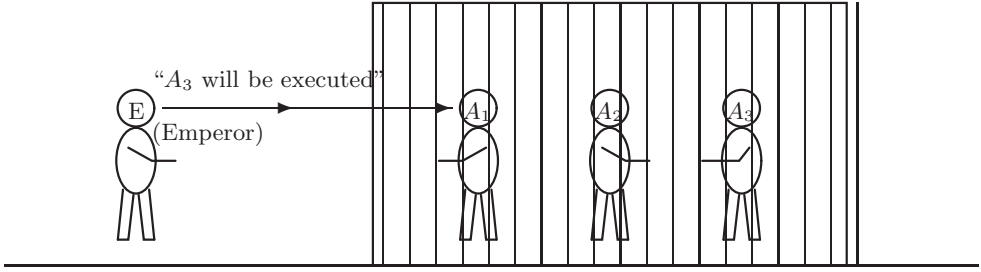
\unitlength=0.35mm
\begin{picture}(500,130)
\thicklines
\put(20,0)
{{{
\put(70,20)
{
\put(0,60){\circle{14}}
\put(0,40){\ellipse{15}{25}}
\path(-3,45)(6,40)(15,40)
\put(-3,56){\footnotesize E}
\path(-7,10)(-5,29)
\path(5,29)(7,10)
\path(-7,10)(-3,10)(-1,27)
\path(1,27)(3,10)(7,10)
}
\put(200,20)
{{
{
\put(0,60){\circle{14}}
\put(0,40){\ellipse{15}{25}}
\path(3,45)(-6,40)(-15,40)
\put(-6,56){\footnotesize $A_1$}
\path(-7,10)(-5,29)
\path(5,29)(7,10)
\path(-7,10)(-3,10)(-1,27)
\path(1,27)(3,10)(7,10)
}
\put(50,0)
{
\put(0,60){\circle{14}}
\put(0,40){\ellipse{15}{25}}
\path(-3,45)(6,40)(15,40)
\put(-6,56){\footnotesize $A_2$}
\path(-7,10)(-5,29)
\path(5,29)(7,10)
\path(-7,10)(-3,10)(-1,27)
\path(1,27)(3,10)(7,10)
}
\put(100,0)
{
\put(0,60){\circle{14}}
\put(0,40){\ellipse{15}{25}}
\path(3,45)(-,40)(-15,40)
\put(-6,56){\footnotesize $A_3$}
\path(-7,10)(-5,29)
\path(5,29)(7,10)
\path(-7,10)(-3,10)(-1,27)
\path(1,27)(3,10)(7,10)
}
}}
\thicklines
\put(20,20){\line(1,0){370}}
\put(160,20){
\path(0,0)(0,100)(180,100)(180,0)
}
\linethickness{0.15mm}
\put(164,20)
{
\multiput(0,0)(10,0){19}{\line(0,1){100}}
}
\put(70,20)
{
\put(10,60){\vector(1,0){50}}
\put(60,60){\vector(1,0){60}}
\put(6,68){\footnotesize \LL $A_3$ will be executed\RR}
\put(6,48){\footnotesize (Emperor)}
}
}}}
\end{picture}

\par
\noindent
{\it Remark 2}.
In Problem 1$'$,
you may choose "Door $A_1$" by various ways.
For example,
you may choose "Door $A_1$" by the method mentioned in
Problem 1$''$ later.

\subsection{
The second answer to
Problems 1$'$ and 2$'$
(
Monty Hall Problem
\textcolor{blue}{[resp. Three prisoners problem]}
)
by Bayes' method
}
\rm

In what follows we study these problems.
Let
$\Omega$ and ${\mathsf O}_1$
be as in Section 3 .
Under the hypothesis
($\sharp_1)$
\textcolor{blue}{[resp. ($\sharp_2)$]},
define the mixed state
$\nu_0$
$({}\in {\cal M}_{+1}^m ({}\Omega{}){})$
such that:
\begin{align}
\nu_0 ({}\{ \omega_1 \}{}) = p_1,
\quad
\nu_0  ({}\{ \omega_2 \}{}) = p_2,
\quad
\nu_0  ({}\{ \omega_3 \}{}) = p_3
\label{eq6}
\end{align}
Thus we have a statistical measurement
${\mathsf M}_{C({}\Omega{})} ({}{\mathsf O}_1, S_{[{}\ast{}]} ({} \nu_0{}))$.
Note that
\begin{itemize}
\item[a)]
{\lq\lq}measured value $1$ is obtained
by the statistical measurement
${\mathsf M}_{C({}\Omega{})} ({}{\mathsf O}_1, 
S_{[{}\ast{}]} ({} \nu_0{}))${\rq\rq}
\\
$
\Leftrightarrow \text{the host says {\lq\lq}Door $A_1$
has a goat{\rq\rq}}
$
\\
\color{blue}
\text{[resp.
$\Leftrightarrow$
the emperor says {\lq\lq}Prisoner $A_1$
will be executed{\rq\rq}
]}
\color{black}
\item[b)]
{\lq\lq}measured value $2$ is obtained
by the statistical measurement
${\mathsf M}_{C({}\Omega{})} ({}{\mathsf O}_1, 
S_{[{}\ast{}]} ({} \nu_0{}))${\rq\rq}
\\
$
\Leftrightarrow \text{the host says {\lq\lq}Door $A_2$
has a goat{\rq\rq}}
$
\\
\color{blue}
\text{[resp.
$\Leftrightarrow$
the emperor says {\lq\lq}Prisoner $A_2$
will be executed{\rq\rq}
]}
\color{black}
\item[c)]
{\lq\lq}measured value $3$ is obtained
by the statistical measurement
${\mathsf M}_{C({}\Omega{})} ({}{\mathsf O}_1, 
S_{[{}\ast{}]} ({} \nu_0{}))${\rq\rq}
\\
$
\Leftrightarrow
$
the host says {\lq\lq}Door $A_3$
has a goat{\rq\rq}
\\
\color{blue}
\text{[resp.
$\Leftrightarrow$
the emperor says {\lq\lq}Prisoner $A_3$
will be executed{\rq\rq}
]}
\color{black}
\end{itemize}
\par
\noindent
Here, assume that,
by the
statistical
measurement
${\mathsf M}_{C({}\Omega{})} ({}{\mathsf O}_1, S_{[{}\ast{}]} ({}\nu_0{}))$,
you obtain a measured value $3$,
which corresponds to
the fact that
the host said
{\lq\lq}Door $A_3$ has a goat{\rq\rq}$\!\!\!.\;$
\color{blue}{[resp.
the emperor said that
Prisoner $A_3$ is to be executed
]},
\color{black}
Then,
Theorem 2
(Bayes' method)
says that
the posterior state $\nu_{\rm post}$
$({}\in {\cal M}_{+1}^m ({}\Omega{}){})$
is given by
\begin{align}
\nu_{\rm post}
= \frac{F_1(\{3\}) \times \nu_0}
{\bigl\langle \nu_0, F_1(\{3\})
\bigr\rangle}.
\label{eq7}
\end{align}
That is,
\begin{align}
&
\nu_{\rm post} ({}\{ \omega_1 \}{})= \frac{\frac{p_1}{2}}{ \frac{p_1}{2} + p_2 },
\quad
\nu_{\rm post} ({}\{ \omega_2 \}{})= \frac{p_2}{ \frac{p_1}{2} + p_2 },
\quad
\nu_{\rm post} ({}\{ \omega_3 \}{}) =  0.
\label{eq8}
\end{align}
Then,
\begin{itemize}
\item[(I1)]
In Problem 1$'$,
$$
\begin{cases}
\mbox{
if 
$\nu_{\rm post} ({}\{ \omega_1 \}{})$
$<$
$\nu_{\rm post} ({}\{ \omega_2 \}{})$
(i.e., $p_1 < 2 p_2 $),
you should pick
Door $A_2$}
\\
\mbox{
if 
$\nu_{\rm post} ({}\{ \omega_1 \}{})$
$=$
$\nu_{\rm post} ({}\{ \omega_2 \}{})$
(i.e., $p_1 < 2 p_2 $),
you may pick
Doors $A_1$ or $A_2$}
\\
\mbox{
if 
$\nu_{\rm post} ({}\{ \omega_1 \}{})$
$>$
$\nu_{\rm post} ({}\{ \omega_2 \}{})$
(i.e., $p_1 < 2 p_2 $),
you should not pick
Door $A_2$}
\end{cases}
$$
\rm
\color{blue}
\item[(I2)]
\color{blue}
In Problem 2$'$,
$$
\begin{cases}
\mbox{
if $
\nu_{0} (\{\omega_1\})
< \nu_{\roman post} (\{\omega_1\})$
(i.e., $p_1  < 1- 2 p_2$),
the prisoner $A_1$'s happiness increases
}
\\
\mbox{
if $
\nu_{0} (\{\omega_1\})
= \nu_{\roman post} (\{\omega_1\})$
(i.e., $p_1  = 1- 2 p_2$),
the prisoner $A_1$'s happiness is invariant
}
\\
\mbox{
if $
\nu_{0} (\{\omega_1\})
> \nu_{\roman post} (\{\omega_1\})$
(i.e., $p_1 > 1- 2 p_2$),
the prisoner $A_1$'s happiness decreases
}
\\
\end{cases}
$$
\rm
\color{black}
\end{itemize}

\par
\par
\noindent
\vskip0.3cm


\section{
The Principle of Equal Probability
}

In this section,
according to
\cite{Ishi2, Ishi4, Ishi9}
we prepare
Theorem 3 (the principle of equal probability),
i.e.,
\begin{itemize}
\item[(J)]
unless we have sufficient reason to regard one possible case
as more probable than another,
we treat them as equally probable.
\end{itemize}
This theorem will be used in the following section.



\vskip0.3cm
%

\par
\noindent
\par
\par
Put
$\Omega = \{ \omega_1 , \omega_2 , \omega_3, \ldots , \omega_n \}$
with the discrete topology.
And consider any observable
${\mathsf O}_1
\equiv (X, {\cal F},  {F}_1 )$
in $C(\Omega)$.

Define the bijection
$\phi_1: \Omega \to \Omega$
such that
\begin{align*}
\phi_1 ( \omega_j ) = 
\cases
\omega_{j+1} & \quad (j \not= n)
\\
\omega_{1} & \quad (j=n)
\endcases
\end{align*}
%
%
and define the observable
${\mathsf O}_k
\equiv (X, {\cal F},  {F}_k )$
in $C(\Omega)$
such that
\begin{align*}
&
[F_k(\Xi)](\omega)
=
[F_1(\Xi)](\phi_{k-1}(\omega))
\\
&
\quad
(\forall \omega \in \Omega, k=1,2,...,n )
\end{align*}
where 
$\phi_0 (\omega) =\omega (\forall \in \Omega )$
and
$\phi_{k} (\omega) = \phi_1 (\phi_{k-1} (\omega ))$
$(\forall \omega \in \Omega, k=1,2,...,n )$.

Let
$p_k (k=1,...,n)$
be a non-negative real number such that
$\sum_{k=1}^n p_k =1$.

\begin{itemize}
\item[(K)]
For example, fix a state $\delta_{\omega_m}$
$(m=1,2,...,n)$.
And,
by the cast of the ( distorted ) dice,
you
choose an observable
$
{\mathsf O}_k
\equiv (X, {\cal F},{ F_k} )
$
with probability
$p_k$.
And further,
you take a measurement
$
{\mathsf M}_{C({}{\Omega})}( {\mathsf O}_k
:= (X, {\cal F},  {F}_k ), 
S_{[\delta_{\omega_m} ]}
)
$.
\end{itemize}
Here, we can easily see that
the probability that a measured value obtained by the
measurement (K)
belongs to
$\Xi
(\in {\cal F})$
is given by
\begin{align}
\sum_{k=1}^n p_k \langle F_k (\Xi) , \delta_{\omega_m} \rangle
\big(
=
\sum_{k=1}^n p_k [ F_k (\Xi)] (\omega_m)
\big)
\label{eq9}
\end{align}
which is equal to
$\langle F_1 (\Xi) , \sum_{k=1}^n p_k \delta_{\phi_{k-1} (\omega_m)} \rangle$.
This implies that
the measurement (K)
is equivalent to
a statistical measurement
$
{\mathsf M}_{C({}{\Omega})}( {\mathsf O}_1
:= (X, {\cal F}, {F}_1 ), 
$
$
S_{[\delta_{\omega_m}]}
(
\sum_{k=1}^n p_k \delta_{\phi_{k-1} (\omega_m)} 
)
)
$.
Note that
the (\ref{eq9}) depends on
the state $\delta_m$.
Thus, we can not calculate the (\ref{eq9}) such as the (\ref{eq8}).

However,
if it holds that
$p_k = 1/n$
$(k=1,...,n)$,
we see that
$
\frac{1}{n} \sum_{k=1}^n  \delta_{\phi_{k-1} (\omega_m) }
$
is independent of the choice of the state
$\delta_{\omega_m}$.
Thus, putting
$
\frac{1}{n} \sum_{k=1}^n  \delta_{\phi_{k-1} (\omega_m) }
$
$=$
$\nu_e$,
we see that
the measurement (K)
is equivalent to
the statistical measurement
$
{\mathsf M}_{C({}{\Omega})}( {\mathsf O}_1,
$
$
S_{[\delta_{\omega_m}]}
(
\nu_e
)
)
$,
which is also equivalent to
$
{\mathsf M}_{C({}{\Omega})}( {\mathsf O}_1,
$
$
S_{[\ast ]}
(
\nu_e
)
)
$
(from the formula (\ref{eq2}) in Remark 1).

Thus, under the above notation,
we have
the following theorem, which realizes the spirit (J).
\par
\noindent
{\bf Theorem 3}
[
The principle of equal probability
(i.e.,
the equal probability of selection)
].
If
$p_k = 1/n $
$(k=1,...,n)$,
the measurement (K)
is independent of the choice of the state $\delta_m$.
Hence, the (K) is equivalent to
a statistical measurement
$
{\mathsf M}_{C({}{\Omega})}( {\mathsf O}_1
:= (X, {\cal F}, {F}_1 ), 
$
$
S_{[\ast]}
(
\nu_e
)
)
$.
\par
\vskip0.3cm
\par
It should be noted that
the principle of equal probability
is not
"principle"
but
"theorem"
in measurement theory.

%

\par
\noindent
{\it Remark 3}.
In the above argument, we consider the set
$B'=\{\phi_k \;|\;k=1,2,...,n \}$.
However,
it may be more natural to consider the set $B=\{\phi \;|\;
\mbox{$\phi: \Omega \to \Omega$ is a bijection}
\}$.
\par
\noindent
\section{
The Third Answer to Monty Hall Problem
\textcolor{blue}{[resp. Three prisoners problem]}
by the principle of equal probability}
\par

\subsection{
Problems 1$''$ and 2$''$
(
Monty Hall Problem
\textcolor{blue}{[resp. Three prisoners problem]}
)
}

\par
\noindent
{\bf Problem 1$''$}
\rm
[{}Monty Hall problem; you cast the dice].
{
Suppose you are on a game show, and you are given
the choice of three doors
(i.e., \LL Door $A_1$\RR$\!\!\!,\;$ \LL Door $A_2$\RR$\!\!\!,\;$ \LL Door $A_3$\RR$\!\!)$.
Behind one door is a car, behind the others, goats.
You do not know what's behind the doors.
Thus,
\begin{itemize}
\rm
\item[{}{($\sharp_1$)}]
\it
you select Door $A_1$
by the cast of the fair dice.
That is,
you say "Door $A_1$"
with probability
1/3.
\end{itemize}
\par
\noindent
The host,
who knows what's behind the doors, opens another door,
say \LL Door $A_3$\RR$\!\!\!,\;$ which has a goat.
He says to you,
\LL Do you want to pick Door $A_2$?\RR$\;\;$
Is it to your advantage to switch your choice of doors?
\par
\noindent
\vskip0.3cm
\par
\noindent
\unitlength=0.26mm
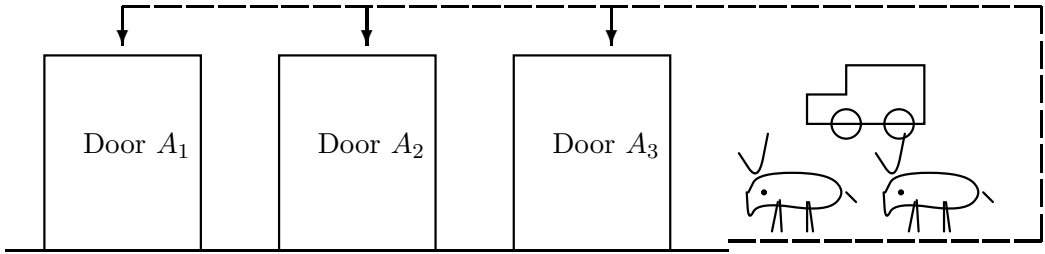
\begin{picture}(500,150)
\thicklines
\put(430,55)
{{
\drawline[-15](-40,-30)(120,-30)(120,90)(-350,90)
\put(-350,90){\vector(0,-1){20}}
\put(-225,90){\vector(0,-1){20}}
\put(-100,90){\vector(0,-1){20}}
\path(0,30)(60,30)(60,60)(20,60)(20,45)(0,45)(0,30)
\put(20,30){\circle{15}}
\put(47,30){\circle{15}}
}}
\put(400,20)
{{
\spline(0,30)(5,40)(40,40)(50,30)(40,20)(5,25)(0,15)(-1,30)
\spline(-5,50)(5,35)(10,60)
\path(15,25)(12,10)
\path(16,26)(17,10)
\path(30,25)(30,10)
\path(31,25)(33,10)
\put(8,30){\circle*{2}}
\path(50,30)(55,25)
}}
\put(470,20)
{{
\spline(0,30)(5,40)(40,40)(50,30)(40,20)(5,25)(0,15)(-1,30)
\spline(-5,50)(5,35)(10,60)
\path(15,25)(12,10)
\path(16,26)(17,10)
\path(30,25)(30,10)
\path(31,25)(33,10)
\put(8,30){\circle*{2}}
\path(50,30)(55,25)
}}

\thicklines
\put(20,20){\line(1,0){370}}
\put(40,20){
\path(0,0)(0,100)(80,100)(80,0)
\put(20,50){Door $A_1$}
}
\put(160,20){
\path(0,0)(0,100)(80,100)(80,0)
\put(20,50){Door $A_2$}
}
\put(280,20){
\path(0,0)(0,100)(80,100)(80,0)
\put(20,50){Door $A_3$}
}
\end{picture}

\par
\noindent
{\bf Problem 2$''$}
\rm
[{}Three prisoners problem; the prisoners cast the dice].
Three prisoners, $A_1$, $A_2$, and $A_3$ were in jail.
They knew that one of them was to be set free and
the other two were to be executed.
They did not know who was the one to be spared,
but the emperor did know.
%
%
Since three prisoners wanted to ask the emperor,
\begin{itemize}
\rm
\item[{}{($\sharp_2$)}]
\it
the questioner was decided by the fair die throw. 
And Prisoner $A_1$ was selected
with probability
$1/3$
\end{itemize}
Then, $A_1$ said to the emperor, 
{\LL}I already know that at least one the other two prisoners will be executed, so if you tell me the name of one who will be executed, you won't have given me any information about my own execution\RR.$\;\;$
After some thinking, the emperor said,
\LL $A_3$ will be executed.\RR$\;\;$
Thereupon $A_1$ felt happier because
his chance had increased from $\frac{1}{3(= {\roman Num}[\{A_1,A_2,A_3 \}])}$ to 
$\frac{1}{2(= {\roman Num}[\{ A_1,A_2 \}])}$.
This prisoner $A_1$'s happiness may or may not be reasonable?
\par
\noindent
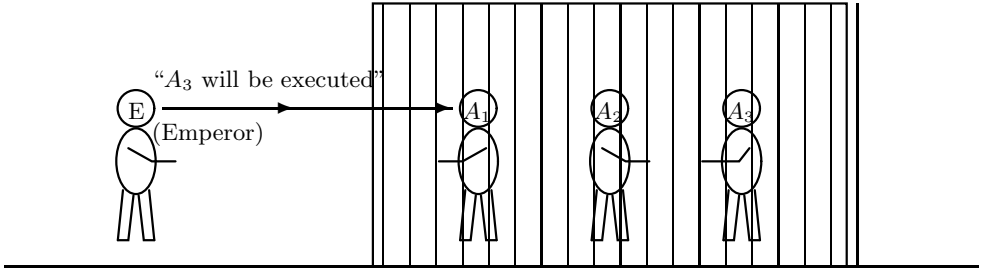
\unitlength=0.35mm
\begin{picture}(500,130)
\thicklines
\put(20,0)
{{{
\put(70,20)
{
\put(0,60){\circle{14}}
\put(0,40){\ellipse{15}{25}}
\path(-3,45)(6,40)(15,40)
\put(-3,56){\footnotesize E}
\path(-7,10)(-5,29)
\path(5,29)(7,10)
\path(-7,10)(-3,10)(-1,27)
\path(1,27)(3,10)(7,10)
}
\put(200,20)
{{
{
\put(0,60){\circle{14}}
\put(0,40){\ellipse{15}{25}}
\path(3,45)(-6,40)(-15,40)
\put(-6,56){\footnotesize $A_1$}
\path(-7,10)(-5,29)
\path(5,29)(7,10)
\path(-7,10)(-3,10)(-1,27)
\path(1,27)(3,10)(7,10)
}
\put(50,0)
{
\put(0,60){\circle{14}}
\put(0,40){\ellipse{15}{25}}
\path(-3,45)(6,40)(15,40)
\put(-6,56){\footnotesize $A_2$}
\path(-7,10)(-5,29)
\path(5,29)(7,10)
\path(-7,10)(-3,10)(-1,27)
\path(1,27)(3,10)(7,10)
}
\put(100,0)
{
\put(0,60){\circle{14}}
\put(0,40){\ellipse{15}{25}}
\path(3,45)(-,40)(-15,40)
\put(-6,56){\footnotesize $A_3$}
\path(-7,10)(-5,29)
\path(5,29)(7,10)
\path(-7,10)(-3,10)(-1,27)
\path(1,27)(3,10)(7,10)
}
}}
\thicklines
\put(20,20){\line(1,0){370}}
\put(160,20){
\path(0,0)(0,100)(180,100)(180,0)
}
\linethickness{0.15mm}
\put(164,20)
{
\multiput(0,0)(10,0){19}{\line(0,1){100}}
}
\put(70,20)
{
\put(10,60){\vector(1,0){50}}
\put(60,60){\vector(1,0){60}}
\put(6,68){\footnotesize \LL $A_3$ will be executed\RR}
\put(6,48){\footnotesize (Emperor)}
}
}}}
\end{picture}

\par
\noindent
\bf
Answer:
\rm
By Theorem 3 (The principle of equal probability),
the above Problems 1$''$ and 2$''$
is respectively the same as Problems 1$'$ and 2$'$
in the case that $p_1=p_2=p_3=1/3$.
Then, the formulas (\ref{eq6}) and (\ref{eq8}) say that
\begin{itemize}
\item[(L1)]
In Problem 1$''$,
since
$\nu_{\rm post} ({}\{ \omega_1 \}{})=1/3$
$<$
$2/3 =\nu_{\rm post} ({}\{ \omega_2 \}{})$,
you should pick
Door $A_2$.
\rm
\color{blue}
\item[(L2)]
\color{blue}
In Problem 2$''$,
since
$
\nu_{0} (\{\omega_1\})=1/3=
\nu_{\roman post} (\{\omega_1\})$,
the prisoner $A_1$'s happiness is invariant.
\rm
\color{black}
\end{itemize}


\rm
\noindent

\par
\noindent

\section{
Conclusions
}

Although main idea is due to
refs.
\cite{Ishi3, Ishi9},
in this paper
we simultaneously discussed the Monty Hall problem
and the three prisoners problem
in terms of quantum language.
That is, we gave three answers,
i.e.,
\begin{itemize}
\item[(M1)]
the first answer (due to Fisher's method) in Section 3,
\item[(M2)]
the second answer (due to Bayes' method)  in Section 4,
\item[(M3)]
the third answer  (due to Theorem 3(the principle of equal probability)) in Section 6
\end{itemize}

We of course believe that our proposal is the final solutions of the two problems. 
It should be noted that
both the Monty Hall problem
and the three prisoners problem are never elementary,
and they
can not be solved without the deep understanding
of "probability" and "dualism (G)".
Thus in this paper, we answered the question:
\begin{center}
"Why have philosophers continued to stick to these problems?"
\end{center}

We hope that
our assertion will be examined from various view points.

\rm
\par
{
\small

\normalsize
}

\end{document}